\documentclass[prb,reprint,longbibliography]{revtex4-1} 

\usepackage{amsmath} 
\usepackage{amsfonts} 
\usepackage{graphicx} 
\usepackage{textgreek}

\usepackage{chngcntr}
\counterwithin{paragraph}{subsection} 

\begin{document}

\title{User-friendly, theory-based web applet for rapidly predicting structure and thermodynamics of complex fluids}

\author{Theodore R. Popp III}
\affiliation{McKetta Department of Chemical Engineering, The University of Texas at Austin, Austin, TX 78712}

\author{Kyle B. Hollingshead}
\affiliation{McKetta Department of Chemical Engineering, The University of Texas at Austin, Austin, TX 78712}

\author{Thomas M. Truskett}
\email{truskett@che.utexas.edu}
\affiliation{McKetta Department of Chemical Engineering, The University of Texas at Austin, Austin, TX 78712}

\date{\today}

\begin{abstract}
Based on a recently introduced analytical strategy [Hollingshead et al., {\em J. Chem. Phys.} {\bf 139}, 161102 (2013)], we present a web applet that can quickly and semi-quantitatively estimate the equilibrium radial distribution function and related thermodynamic properties of a fluid from knowledge of its pair interaction.
We present a detailed description of the applet's features and intended workflow, followed by a description of how the applet can be used to illustrate two (of many possible) concepts of interest for introductory statistical mechanics courses: the transition from ideal gas-like behavior to correlated-liquid behavior with increasing density and the tradeoff between dominant length scales with changing temperature in a system with ramp-shaped repulsions. 
The latter type of interaction qualitatively captures distinctive thermodynamic properties of liquid water because its energetic bias toward locally open structures mimics that of water's hydrogen-bond network.
\end{abstract}

\maketitle

\section{Introduction}
Statistical mechanics provides quantitative links between a fluid's interparticle interactions and its resulting equilibrium structure and thermodynamic properties. 
However, particularly for dense systems or systems with complex interactions, it can be challenging to find ways for students to explore these relationships within the framework of a university course due to the prohibitive amount of time, expertise (either computational or experimental), and/or resources required to, e.g., numerically solve the Ornstein-Zernike relation with an appropriate closure,\cite{thysimpliq,TangLu:FMSA} construct a molecular simulation to extract relevant equilibrium data,\cite{rapaport2004art,allen1989computer,Mulero:RDFSoftware} or carry out relevant measurements in a laboratory.\cite{Younge:RDFExperiment,Heinen:Yukawa}
As a result, for students to become familiar with the relevant concepts, additional tools are required that help them to overcome these technical hurdles.

Here, we present a web-based applet that helps to accomplish this through use of a new analytic integral equation-based method for equilibrium fluids in three dimensions.\cite{Hollingshead:DiscretizedFMSA}  
The applet provides rapid and semi-quantitative graphical predictions of structural and thermodynamic quantities from knowledge of the pair interaction and parameters that describe the thermodynamic state (i.e., density and temperature).
Apart from awareness of a few practical constraints, detailed knowledge of the internal calculations is not required to make productive use of the applet as a pedagogical tool or as an experimental guide.
Because of its efficiency and accessible layout, students are empowered to interactively experiment with a fluid's pair potential or its thermodynamic state and extract meaningful relationships and trends.\cite{Casperson:VisualizationTeaching,Tobochnik:StatPhysTeach,Wieman:PhetSoftware,Laverty:Teaching,Buffler:ModelBasedTeaching} 


\section{Internal Calculations}
The applet accepts as inputs a pairwise potential \(\varphi(r)\) as a function of interparticle separation \(r\), the temperature \(T\), and the number density \(\rho\), and it approximately calculates the corresponding unique radial distribution function (RDF)\cite{henderson:uniqueg} as well as other related thermodynamic quantities. 
The applet requires that the interactions be isotropic, consisting of a hard core of diameter \(\sigma\) plus an arbitrary short-ranged contribution \(\varepsilon \phi(r)\), where \(\varepsilon\) is a characteristic energy scale, that decays to zero by \(r = 2\sigma\),
\begin{equation}
\label{eq:potential}
\frac{\varphi(r)}{\varepsilon} = \left\{ \begin{array}{ll}
\infty & ~~r < \sigma ,\\
\phi(r) & ~~\sigma \leq r \leq 2\sigma ,\\ 
0 & ~~r > 2\sigma .
\end{array} \right.
\end{equation}
By choosing different functions for \(\phi(r)\), this generic form encompasses many different types of effective model interactions routinely used to describe the thermodynamics and structure of complex fluids.\cite{Hollingshead:DiscretizedFMSA} Two possible choices--a bare hard-sphere potential, for which \(\phi(r)=0\), that models excluded-volume interactions in fluids and a repulsive ramp potential, for which \(\phi(r)=2-r/\sigma\), that qualitatively captures some distinctive properties of liquid water--are discussed explicitly in this article. Other possible model interactions include, but are not limited to, Yukawa potentials that model screened electrostatic interactions in colloidal suspensions and dusty plasmas\cite{davoudi:yukawa, cochran:yukawa, Heinen:Yukawa} and the Asakura-Oosawa potential\cite{asakura:asakuraoosawa, roth:asakuraoosawa} that models polymer-mediated depletion interactions between suspended colloids.

\subsection{Radial Distribution Function}
To enable the desired predictions,\cite{Hollingshead:DiscretizedFMSA} the applet first decomposes the continuous potential interaction into a ``terraced'' representation of \(M=100\) equally-spaced discrete steps, each with an outer range 
\begin{equation}
\frac{\lambda_i}{\sigma}= 1+\frac{i}{M}
\label{eq:lambdai}
\end{equation}
and a constant energy 
\begin{equation}
\frac{\varphi_i}{\varepsilon}=\left(\lambda_{i}-\lambda_{i-1}\right)^{-1}\int_{\lambda_{i-1}}^{\lambda_i}\phi(r)\,dr.
\label{eq:phi_i}
\end{equation}
The integration in Eq.~(\ref{eq:phi_i}), and other integrations for the applet are carried out via the trapezoidal rule.

A terraced potential yields a jagged or ``sawtoothed'' RDF, \(g_{\text{ST}}\left(r\right)\), which is computed via an extension\cite{Hollingshead:DiscretizedFMSA} of the simple exponential first-order mean spherical approximation.\cite{Hlushak:ModifiedFMSA} 
Then, to arrive at a continuous RDF prediction that corresponds to the original continuous potential, the ``teeth'' are smoothed by computing a series of linear corrections to \(g_{\text{ST}}(r)\) such that adjacent pieces of the smoothed RDF, \(g(r)\), have equal values at each intersection, i.e. \(g(\lambda_i^-)=g(\lambda_i^+)\), where the superscripts \(^-\) or \(^+\) indicate limiting values approaching each \(\lambda_i\) from the left or right, respectively. 
(See Ref.~\onlinecite{Hollingshead:DiscretizedFMSA} for details). 

\subsection{\label{sec:thermo} Thermodynamic Properties}
The applet calculates several thermodynamic properties that are directly accessible via the pair potential and the RDF.
The internal energy per particle \(u\)
is\cite{thysimpliq} 
\begin{equation}
u = \frac{3 k_{\text B} T}{2} + 2\pi\rho\int_0^\infty{\varphi(r)\,g(r)r^2\,dr},
\end{equation}
where \(k_\text{B}\) is the Boltzmann constant, \(T\) is temperature, \(\rho=N/V\), \(N\) is total number of particles, and \(V\) is volume.
The compressibility factor~\(Z\) is\cite{bannerman:pressure}
\begin{equation}
Z = \frac{\beta P}{\rho} = 1 + \frac{2\pi\rho}{3}\sum_{i=0}^{M}{\lambda_i}^3\left[g_\text{ST}(\lambda_i^+)-g_\text{ST}(\lambda_i^-)\right],
\end{equation}
where \(\beta=(k_\text{B} T)^{-1}\), \(P\) is the pressure, and \(\lambda_i\) is given by Eq.~(\ref{eq:lambdai}). 
Note that the excess Helmholtz free energy of the fluid (and other properties of interest through standard thermodynamics relations) can subsequently be obtained from knowledge of the density and temperature dependence of \(Z\), i.e., the equation of state.\cite{thysimpliq}
The two-body contribution to molar excess entropy~\(s^{(2)}\) is also directly computable from the RDF,\cite{carmer:s2opt,pond:s2gauss,krekelberg:s2}
\begin{equation}
\frac{s^{(2)}}{k_\text{B}}=-2\pi\rho\int_0^\infty \left[g(r)\ln g(r)-g(r)+1\right]r^2\,dr .
\end{equation}
This last quantity--the entropy cost of pair correlations (relative to a structure-free ideal gas)--is of interest because it is known to correlate with dynamic properties (e.g., self diffusivity) in a wide class of fluid systems.\cite{carmer:s2opt,pond:s2gauss,krekelberg:s2,rosenfeld:sexscaling1,rosenfeld:sexscaling2,dzugutov:sexscaling}.

\section{Using the Applet}
The applet is written in Java using the Swing library,\cite{[{}][{. Available online at \textless\url{http://docs.oracle.com/javase/tutorial/uiswing/index.html}\textgreater.}]swinglib} which ensures portability across different operating systems and allows the applet to be embedded in a web page. 
Graphs are created with the JFreeChart library\cite{[][{. Available online at \textless\url{http://www.jfree.org/jfreechart/}\textgreater.}]jfreechart} to allow for easy visualization, manipulation, and analysis of series data. 
A system--comprising the pair potential, the thermodynamic state (\(k_{\text B} T/\varepsilon\) and \(\rho \sigma^3\)), and the calculation parameters--can be saved to a file on the user's computer and reloaded later within the applet. All numerical data can also be exported as tab-separated value (\texttt{.tsv}) text files.

\subsection{System Information}
\begin{figure}[t]
\centering
\includegraphics[width=8.2cm]{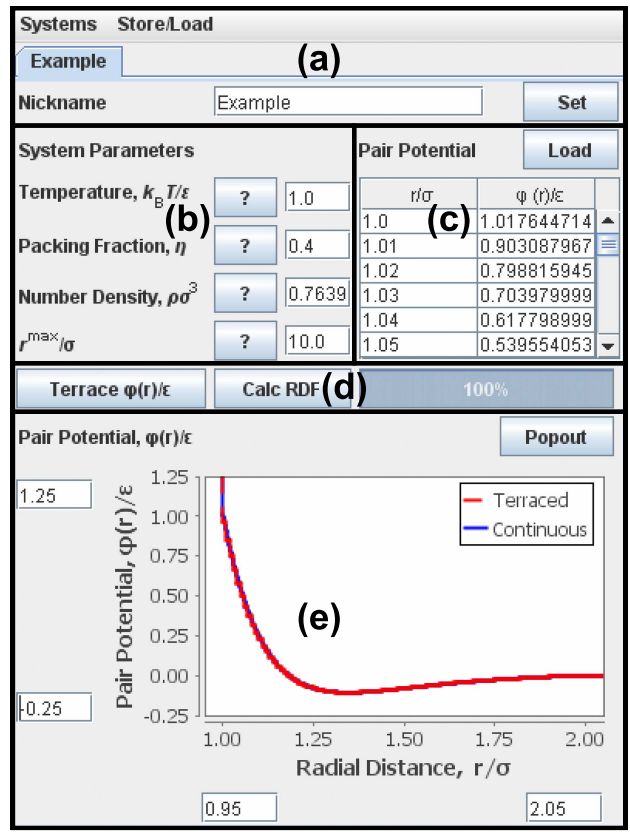}
\caption{The system window in the applet for a selected example. 
(a) Systems can be loaded from and saved to local files; multiple system tabs can be present simultaneously and compared. Each system can be named descriptively. 
(b) Input parameters include the dimensionless temperature \(k_\text{B}T/\varepsilon\), the number density \(\rho\sigma^3\) or packing fraction \(\eta=\rho\pi\sigma^3/6\), and the upper range of the calculation, \(r^\text{max}/\sigma\). 
(c) Each system's short-ranged contribution to the interaction potential can be input as a series of \(\left[r/\sigma,\, \phi(r)\right]\) points, or loaded from a \texttt{.csv} file. 
(d) The terracing and RDF calculations are triggered with buttons, and calculation progress is displayed by the progress bar.
(e) Both the continuous and terraced representations (see text) of the interaction potential are plotted for inspection.}
\label{fig:left}
\end{figure}

Half of the interface is dedicated to receiving user input and displaying information about individual systems (see Fig.~\ref{fig:left}). 
This half of the interface features five sections: (a) controls for opening, closing, saving, or loading systems; (b) user input parameters; (c) a tabular pair potential; (d) buttons to trigger calculations; and (e) plots of the specified potential and its terraced representation.

\paragraph{Opening, closing, and naming systems.}
Each system has a nickname which appears throughout the applet. The applet begins with a single empty system with the default nickname ``System 1.'' The user can provide a new name in the \texttt{Nickname} field, then press \texttt{Set}. When multiple systems are open, the tabs at the top of the panel can be used to switch between the systems.  

Through the \texttt{Systems} menu, the user can create additional empty systems or close the currently focused system tab. Through the \texttt{Store/Load} menu, the user can store the current system in a local file or import a previously saved system. 

\paragraph{Input parameters.}
The user must specify the system's dimensionless temperature \(k_\text{B}T/\varepsilon\), either the number density \(\rho\sigma^3\) or packing fraction \(\eta=\rho\pi\sigma^3/6\), and the range of the calculation, \(r^\text{max}/\sigma\). Care should be taken to ensure that oscillations in the RDF have decayed before \(r^\text{max}/\sigma\). The theoretical approach that the applet relies upon,\cite{Hollingshead:DiscretizedFMSA} similar to most theories of simple liquids,\cite{thysimpliq} loses accuracy near a critical point or in systems with very high density (e.g., \(\rho\sigma^3 \gtrsim 1\)) or very low temperature (e.g., \(k_\text{B}T/\varepsilon \lesssim 0.05\)), with the details depending on the chosen interaction. For most state points away from the critical point of the fluid, the default choice of \(r^\text{max}/\sigma=10\) is conservative. When either \(\eta\) or \(\rho\sigma^3\) is changed, the other field updates automatically. 

\paragraph{Interparticle pair potential.}
The short-ranged addition to the pair potential \(\phi(r)\) can be provided either by editing a table within the applet or by loading the data from an external \texttt{.csv} file. For simple pair potentials constructed from line segments, like a ramp or Jagla potential,\cite{Jagla:ramp} it is sufficient to specify only the end points of each segment. For more complex interactions, however, it is often more convenient  to prepare the potential in a separate file using, e.g., a spreadsheet editing program, then press the \texttt{Load} button in the applet to import the pair potential data.

\paragraph{Performing calculations.}
Once inputs have been provided, the user may click either \texttt{Terrace {\textphi}(r)/\textepsilon} or \texttt{Calc RDF} to view the terraced pair potential or begin calculation of the radial distribution function, respectively. The user may proceed directly from providing inputs to calculating the RDF, but it is recommended that the terraced potential be generated and inspected before beginning the more intensive RDF calculations. The bar to the right of these buttons depicts the progress of the RDF calculation.

\paragraph{Plot of pair potentials.}
The continuous and terraced pair potentials are presented graphically for easy inspection. The continuous curve is updated in real-time as the pair potential is edited; the terraced representation is added when either of the calculation buttons is pressed. By default, the plot shows the full data sets; to focus on a region of interest, the user can click and drag a rectangle, or edit the values in the boxes along the axes to specify an exact window. If desired, the chart can be reproduced in a separate window by pressing the \texttt{Popout} button.
Many more charting options, built into the JFreeChart library, are available by right-clicking the plot. 

\subsection{Comparing Structure and Thermodynamics}
\begin{figure}[t]
\centering
\includegraphics[width=8.2cm]{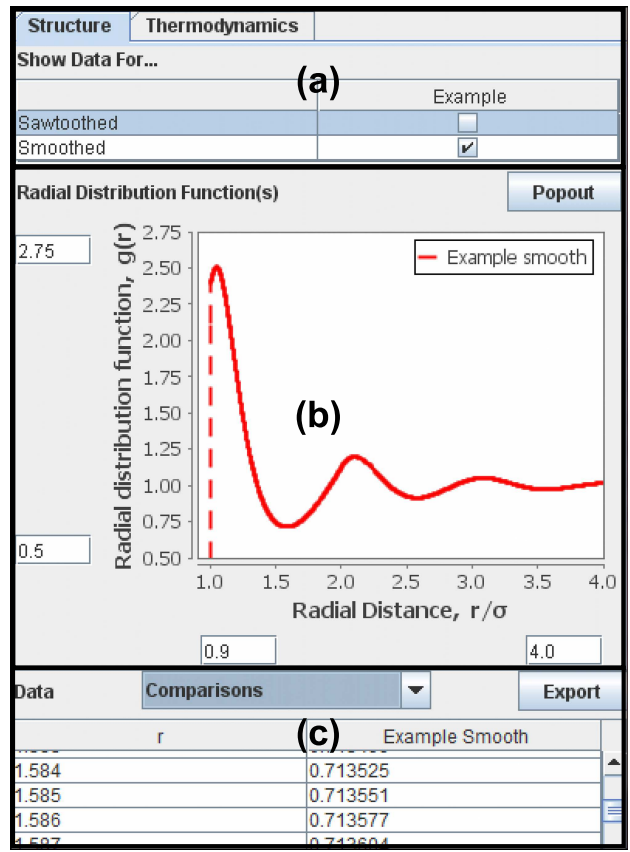}
\caption{The sawtoothed and smoothed radial distribution functions for all systems currently calculated are available for comparison. 
(a) Each curve can be toggled on or off to facilitate comparisons between specific systems. 
The radial distribution functions are both (b) plotted graphically, for visual inspection, and (c) available as data series, to precisely compare specific values. 
The data series can also be exported as a \texttt{.tsv} file.}
\label{fig:right}
\end{figure}

The second half of the applet interface allows the user to view the calculated RDFs and compare them across multiple systems. 
This half contains sections for (a) selecting which systems and system data to compare; (b) plots of the selected RDF predictions; and (c) tabulated numerical data for the selected systems (see Fig.~\ref{fig:right}).

\paragraph{Selecting systems.}
Once a system's RDF calculation is complete, the user can choose to inspect the resulting \(g_\text{ST}(r)\) or \(g(r)\) data by selecting the appropriate checkboxes. 
Multiple data series can be selected simultaneously so the user can compare different systems and analyze, e.g., the impact of the smoothing algorithm or the differences in RDF structure between two systems.

\paragraph{Plot of radial distribution functions.}
The selected radial distribution functions are presented graphically for immediate comparison. 
This plot functions identically to 
the pair potential plot described earlier.

\paragraph{Data tables.}
Numerical intermediate data are available in tabular form for all of the active systems, including continuous and terraced \(\varphi(r)\) representations, and sawtoothed and smoothed RDFs. 
An additional table, labeled \texttt{Comparisons} in the drop-down box, contains all of the data series corresponding to the current state of the RDF plot. 
Any data in these data tables can be copied and pasted; or, the user can press the \texttt{Export} button to save the selected table as a tab-separated value (\texttt{.tsv}) file, which can then be manipulated with a text editing or spreadsheet program.

The \texttt{Thermodynamics} tab, not pictured, contains a table with the thermodynamic properties described in Section \ref{sec:thermo}---average configurational energy, average internal energy, compressibility factor, and two-body excess entropy---calculated for each system. It also features togglable data series and can be used in the same ways as the RDF data tables described above.

\section{Teaching Examples}
This applet offers many pedagogical opportunities to teachers and students of classical statistical mechanics. 
Most simply, it can illustrate the effects of changing temperature, density, or interactions on the fly, e.g. during a lecture.
The applet can also be used to prepare example figures through the use of the plot saving functionality available within the applet, or by exporting the calculated data and plotting in a preferred environment. 
Because of its ability to save and load states, an example ``initial state'' could be prepared for further manipulation during a lecture, or distributed as part of a homework assignment. 
Students are also able to experiment freely by modifying the attractions or repulsions, changing the density or temperature, etc., to develop an intuition for complex fluid phenomena, without needing a simulation suite or more advanced statistical mechanics coding knowledge.

Here we provide examples of how our applet might be used to illustrate two fundamental ideas. 

\subsection{Emerging Coordination Shell Structure with Density}

The hard sphere (HS) fluid---whose particles have no interaction other than a volume exclusion to prevent interparticle overlap, e.g. \(\phi(r)=0\) in Eq.~(\ref{eq:potential})---is a canonical reference model for the structure of dense liquid and colloidal systems, and it is one of the simplest models of a non-ideal gas.
Because the interaction potential is either infinite or zero, its structure is independent of temperature (as are its energies and dynamics, apart from a trivial scaling related to particle velocities).\cite{thysimpliq}
Despite their simplicity, hard sphere fluids (like atomic liquids and particle suspensions) develop nontrivial structure (e.g., interparticle correlations) as density increases. At \(\eta \approx 0.494\), the HS fluid experiences a purely entropy-driven freezing transition to form an FCC crystal.\cite{thysimpliq}

\begin{figure}[t]
\centering
\includegraphics[width=8.2cm]{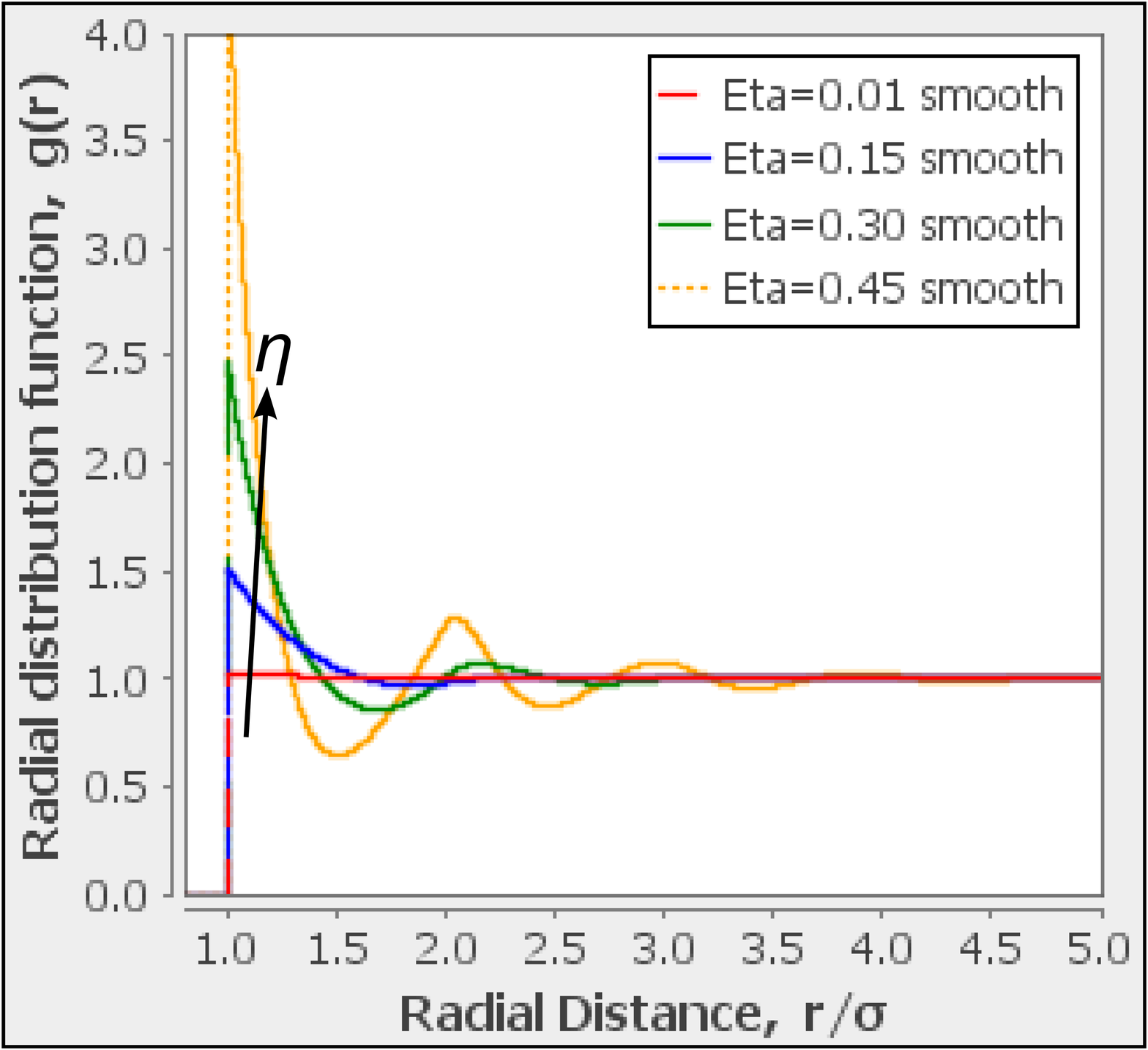}
\caption{The amount of structure increases with increasing packing fraction in a hard sphere fluid; shown are packing fractions \(\eta=0.01\) (red), where almost no correlations are present beyond the hard core; \(\eta=0.15\) (blue); \(\eta=0.30\) (green); and \(\eta=0.45\) (orange), where correlations extend to nearly six particle diameters and a large population of particles are in contact.}
\label{fig:density}
\end{figure}

In Fig.~\ref{fig:density}, we have used the applet to plot the radial distribution functions of HS fluid systems at \(\eta = 0.01\), \(0.15\), \(0.30\), and \(0.45\). 
As the packing fraction is increased, several trends can be readily observed: first, the range of the correlation increases from slightly beyond \(r/\sigma=1\) to nearly \(r/\sigma=5\) as coordination shells of nearest, next-nearest neighbors (and so on) develop; second, the magnitude of the first peak in the radial distribution function increases from \(g(r) \approx 1\) to \(g(r) \approx 5\), indicating that particles are contacting one another with greater and greater frequency; and third, the period of the oscillations (once they are present) shrinks as the coordination shells become more condensed. These structural trends with increasing density, also commonly seen in simple liquids, result in an increased pressure and reduced excess entropy--both of which are readily verifiable in the applet.

\subsection{Temperature Effects in a Two-Length-Scale Fluid}
\begin{figure}[t]
\centering
\includegraphics[width=5.8cm]{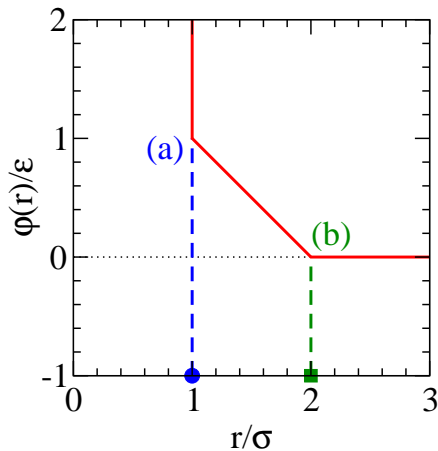}
\caption{The repulsive ramp potential (red solid curve) has two length scales: the hard core diameter (a, blue circle), and the outer edge of the ramp (b, green square). The former is favored at high temperature, while the latter is favored at low temperature.} 
\label{fig:ramppot}
\end{figure}

In liquids more complex than hard spheres, multiple length scales can be present within the pair potential. 
For example, in a repulsive ramp system where 
\begin{equation}
\label{eq:ramp}
\phi(r) = 2 - \frac{r}{\sigma},
\end{equation}
there are relevant length scales at \(r/\sigma=1\), at the edge of the hard core, and at \(r/\sigma=2\), at the outer limit of the interaction (see Fig.~\ref{fig:ramppot}).
For an interaction of this form, one might expect that at high temperatures (\(k_\text{B}T \gg \varepsilon \)), the energy associated with the finite repulsion outside \(r=\sigma\) would be negligible relative to the thermal energy of the system; therefore, the hard core length scale might be most relevant (i.e, the system approaches hard-sphere-like structure). 
Conversely, at low temperatures (\(k_\text{B}T \ll \varepsilon\)), contributions from the finite repulsion would be more significant, leading the \(r=2\sigma\) (more open, low density) length scale to dominate. 

\begin{figure}[t]
\centering
\includegraphics[width=8.2cm]{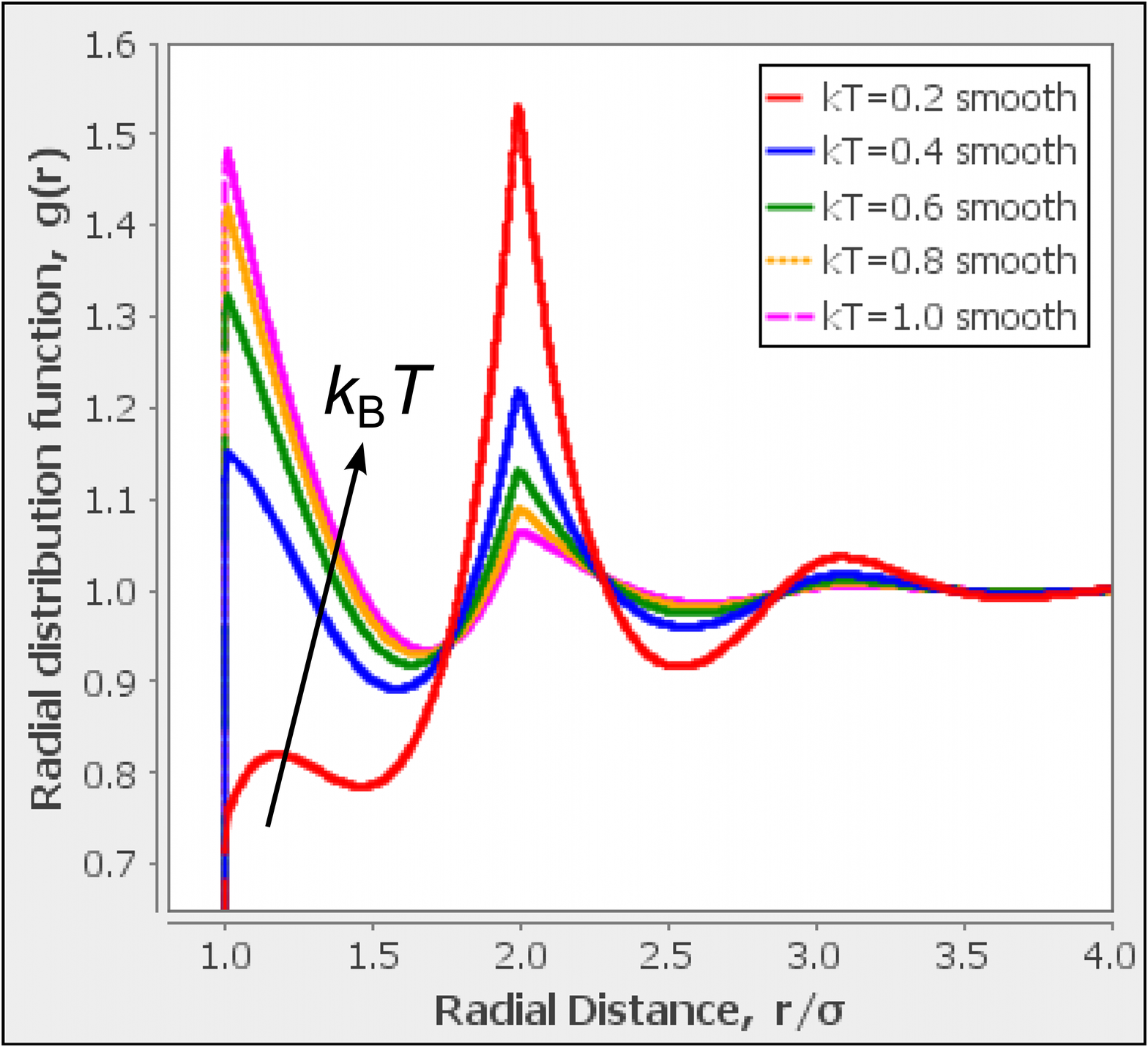}
\caption{Smoothed radial distribution functions of ramp systems (see Eq.~(\ref{eq:ramp})) plotted with the applet, where \(\eta=0.2\) and \(k_\text{B}T=0.2\) (red), 
\(0.4\) (blue), \(0.6\) (green), \(0.8\) (orange), and \(1.0\) (magenta).}
\label{fig:ramps}
\end{figure}

The applet can be used to demonstrate this phenomenon, by simulating the same ramp potential at a series of different temperatures, \(k_\text{B}T/\varepsilon = 0.2\), \(0.4\), \(0.6\), \(0.8\), and \(1.0\) (see Fig.~\ref{fig:ramps}).
At all of these temperatures, clear strong peaks corresponding to the two length scales in the pair potential are present at \(r/\sigma=1\) and \(r/\sigma=2\). However, as the temperature decreases, so too does the significance of the inner peak; below \(k_\text{B}T/\varepsilon = 0.4\), the outer peak is taller. 

These temperature-dependent trends are analogous to those seen in network-forming fluids like liquid water, where the hydrogen-bond network energetically favors low-coordinated, open structures. Due to analogous physics along isobars, at moderate pressures, such structures dominate in water, leading the fluid to exhibit negative thermal expansivity (expansion upon cooling) at low temperature--a thermodynamic property also exhibited by the ramp model.\cite{Jagla:ramp, yan:ramp, errington:ramp}  Similar features occur in other network forming fluids like silica whose interactions energetically favor locally open structures as well.

\section{Conclusion}
This applet provides new opportunities for students and teachers of statistical mechanics to explore and develop a deeper conceptual understanding of the effects of interparticle interactions and the thermodynamic state on the particle-scale equilibrium structure and thermodynamic properties in a fluid system.

The applet is freely available for use or download at \textless\url{http://www.truskettgroup.com/fluidapp/}\textgreater, and its source code is available under the GNU General Public License\cite{[{}][{. License text available online at \textless\url{http://www.gnu.org/copyleft/gpl.html}\textgreater.}]gpl} on GitHub.\cite{[{}][{Available online at \textless\url{https://github.com/TRP3/FluidRDFApp}\textgreater.}]github} 
The authors encourage any interested parties to modify or expand the applet in useful ways. 
We hope to expand its functionality in the future, most immediately by adding options for the use of additional integral equation theory closures in order to treat an even broader variety of possible pair interactions.
We also intend to implement a calculation of the structure factor,
\begin{equation}
S(k)=1+4\pi\rho\int_0^\infty{\frac{\sin(kr)}{kr}\left[g(r)-1\right]r^2\,dr,}
\end{equation}
which is an experimentally accessible quantity that can offer insight into, e.g., freezing transitions via the Hansen-Verlet freezing criterion.\cite{hansen:freezing}

\section{Acknowledgements}
The authors acknowledge support from the Welch Foundation (F-1696) and the National Science Foundation (CBET-1065357).


\begin{thebibliography}{34}%
\makeatletter
\providecommand \@ifxundefined [1]{%
 \@ifx{#1\undefined}
}%
\providecommand \@ifnum [1]{%
 \ifnum #1\expandafter \@firstoftwo
 \else \expandafter \@secondoftwo
 \fi
}%
\providecommand \@ifx [1]{%
 \ifx #1\expandafter \@firstoftwo
 \else \expandafter \@secondoftwo
 \fi
}%
\providecommand \natexlab [1]{#1}%
\providecommand \enquote  [1]{``#1''}%
\providecommand \bibnamefont  [1]{#1}%
\providecommand \bibfnamefont [1]{#1}%
\providecommand \citenamefont [1]{#1}%
\providecommand \href@noop [0]{\@secondoftwo}%
\providecommand \href [0]{\begingroup \@sanitize@url \@href}%
\providecommand \@href[1]{\@@startlink{#1}\@@href}%
\providecommand \@@href[1]{\endgroup#1\@@endlink}%
\providecommand \@sanitize@url [0]{\catcode `\\12\catcode `\$12\catcode
  `\&12\catcode `\#12\catcode `\^12\catcode `\_12\catcode `\%12\relax}%
\providecommand \@@startlink[1]{}%
\providecommand \@@endlink[0]{}%
\providecommand \url  [0]{\begingroup\@sanitize@url \@url }%
\providecommand \@url [1]{\endgroup\@href {#1}{\urlprefix }}%
\providecommand \urlprefix  [0]{URL }%
\providecommand \Eprint [0]{\href }%
\providecommand \doibase [0]{http://dx.doi.org/}%
\providecommand \selectlanguage [0]{\@gobble}%
\providecommand \bibinfo  [0]{\@secondoftwo}%
\providecommand \bibfield  [0]{\@secondoftwo}%
\providecommand \translation [1]{[#1]}%
\providecommand \BibitemOpen [0]{}%
\providecommand \bibitemStop [0]{}%
\providecommand \bibitemNoStop [0]{.\EOS\space}%
\providecommand \EOS [0]{\spacefactor3000\relax}%
\providecommand \BibitemShut  [1]{\csname bibitem#1\endcsname}%
\let\auto@bib@innerbib\@empty
\bibitem [{\citenamefont {Hansen}\ and\ \citenamefont
  {McDonald}(2006)}]{thysimpliq}%
  \BibitemOpen
  \bibfield  {author} {\bibinfo {author} {\bibfnamefont {J.-P.}\ \bibnamefont
  {Hansen}}\ and\ \bibinfo {author} {\bibfnamefont {I.R.}\ \bibnamefont
  {McDonald}},\ }\href {http://books.google.com/books?id=Uhm87WZBnxEC} {\emph
  {\bibinfo {title} {Theory of Simple Liquids}}}\ (\bibinfo  {publisher}
  {Elsevier Science},\ \bibinfo {year} {2006})\BibitemShut {NoStop}%
\bibitem [{\citenamefont {Tang}\ and\ \citenamefont {Lu}(1997)}]{TangLu:FMSA}%
  \BibitemOpen
  \bibfield  {author} {\bibinfo {author} {\bibfnamefont {Yiping}\ \bibnamefont
  {Tang}}\ and\ \bibinfo {author} {\bibfnamefont {Benjamin C.-Y.}\ \bibnamefont
  {Lu}},\ }\bibfield  {title} {\enquote {\bibinfo {title} {Analytical
  representation of the radial distribution function for classical fluids},}\
  }\href {\doibase 10.1080/002689797172697} {\bibfield  {journal} {\bibinfo
  {journal} {Mol. Phys.}\ }\textbf {\bibinfo {volume} {90}},\ \bibinfo {pages}
  {215--224} (\bibinfo {year} {1997})}\BibitemShut {NoStop}%
\bibitem [{\citenamefont {Rapaport}(2004)}]{rapaport2004art}%
  \BibitemOpen
  \bibfield  {author} {\bibinfo {author} {\bibfnamefont {D.C.}\ \bibnamefont
  {Rapaport}},\ }\href {http://books.google.com/books?id=iqDJ2hjqBMEC} {\emph
  {\bibinfo {title} {The Art of Molecular Dynamics Simulation}}}\ (\bibinfo
  {publisher} {Cambridge University Press},\ \bibinfo {year}
  {2004})\BibitemShut {NoStop}%
\bibitem [{\citenamefont {Allen}\ and\ \citenamefont
  {Tildesley}(1989)}]{allen1989computer}%
  \BibitemOpen
  \bibfield  {author} {\bibinfo {author} {\bibfnamefont {M.P.}\ \bibnamefont
  {Allen}}\ and\ \bibinfo {author} {\bibfnamefont {D.J.}\ \bibnamefont
  {Tildesley}},\ }\href {http://books.google.com/books?id=Y0jEngEACAAJ} {\emph
  {\bibinfo {title} {Computer Simulation of Liquids}}},\ Oxford Science Publ\
  (\bibinfo  {publisher} {Clarendon Press},\ \bibinfo {year}
  {1989})\BibitemShut {NoStop}%
\bibitem [{\citenamefont {Mulero}\ \emph {et~al.}(1993)\citenamefont {Mulero},
  \citenamefont {Cuadros},\ and\ \citenamefont
  {P\'{e}rez-Ayala}}]{Mulero:RDFSoftware}%
  \BibitemOpen
  \bibfield  {author} {\bibinfo {author} {\bibfnamefont {A.}~\bibnamefont
  {Mulero}}, \bibinfo {author} {\bibfnamefont {F.}~\bibnamefont {Cuadros}}, \
  and\ \bibinfo {author} {\bibfnamefont {M.}~\bibnamefont {P\'{e}rez-Ayala}},\
  }\bibfield  {title} {\enquote {\bibinfo {title} {Displaying the role of
  repulsive and attractive intermolecular forces in fluids},}\ }\href@noop {}
  {\bibfield  {journal} {\bibinfo  {journal} {American Journal of Physics}\
  }\textbf {\bibinfo {volume} {61}} (\bibinfo {year} {1993})}\BibitemShut
  {NoStop}%
\bibitem [{\citenamefont {Younge}\ \emph {et~al.}(2004)\citenamefont {Younge},
  \citenamefont {Christenson}, \citenamefont {Bohara}, \citenamefont
  {Crnkovic},\ and\ \citenamefont {Saulnier}}]{Younge:RDFExperiment}%
  \BibitemOpen
  \bibfield  {author} {\bibinfo {author} {\bibfnamefont {K.}~\bibnamefont
  {Younge}}, \bibinfo {author} {\bibfnamefont {C.}~\bibnamefont {Christenson}},
  \bibinfo {author} {\bibfnamefont {A.}~\bibnamefont {Bohara}}, \bibinfo
  {author} {\bibfnamefont {J.}~\bibnamefont {Crnkovic}}, \ and\ \bibinfo
  {author} {\bibfnamefont {P.}~\bibnamefont {Saulnier}},\ }\bibfield  {title}
  {\enquote {\bibinfo {title} {{A model system for examining the radial
  distribution function}},}\ }\href {\doibase 10.1119/1.1758228} {\bibfield
  {journal} {\bibinfo  {journal} {Am. J. Phys.}\ }\textbf {\bibinfo {volume}
  {72}},\ \bibinfo {pages} {1247} (\bibinfo {year} {2004})}\BibitemShut
  {NoStop}%
\bibitem [{\citenamefont {Heinen}\ \emph {et~al.}(2011)\citenamefont {Heinen},
  \citenamefont {Holmqvist}, \citenamefont {Banchio},\ and\ \citenamefont
  {N\"{a}gele}}]{Heinen:Yukawa}%
  \BibitemOpen
  \bibfield  {author} {\bibinfo {author} {\bibfnamefont {Marco}\ \bibnamefont
  {Heinen}}, \bibinfo {author} {\bibfnamefont {Peter}\ \bibnamefont
  {Holmqvist}}, \bibinfo {author} {\bibfnamefont {Adolfo~J}\ \bibnamefont
  {Banchio}}, \ and\ \bibinfo {author} {\bibfnamefont {Gerhard}\ \bibnamefont
  {N\"{a}gele}},\ }\bibfield  {title} {\enquote {\bibinfo {title} {{Pair
  structure of the hard-sphere Yukawa fluid: an improved analytic method versus
  simulations, Rogers-Young scheme, and experiment.}}}\ }\href {\doibase
  10.1063/1.3524309} {\bibfield  {journal} {\bibinfo  {journal} {J. Chem.
  Phys.}\ }\textbf {\bibinfo {volume} {134}},\ \bibinfo {pages} {044532}
  (\bibinfo {year} {2011})}\BibitemShut {NoStop}%
\bibitem [{\citenamefont {Hollingshead}\ \emph {et~al.}(2013)\citenamefont
  {Hollingshead}, \citenamefont {Jain},\ and\ \citenamefont
  {Truskett}}]{Hollingshead:DiscretizedFMSA}%
  \BibitemOpen
  \bibfield  {author} {\bibinfo {author} {\bibfnamefont {Kyle~B.}\ \bibnamefont
  {Hollingshead}}, \bibinfo {author} {\bibfnamefont {Avni}\ \bibnamefont
  {Jain}}, \ and\ \bibinfo {author} {\bibfnamefont {Thomas~M.}\ \bibnamefont
  {Truskett}},\ }\bibfield  {title} {\enquote {\bibinfo {title}
  {{Communication: Fine discretization of pair interactions and an approximate
  analytical strategy for predicting equilibrium behavior of complex
  fluids.}}}\ }\href {\doibase 10.1063/1.4826649} {\bibfield  {journal}
  {\bibinfo  {journal} {J. Chem. Phys.}\ }\textbf {\bibinfo {volume} {139}},\
  \bibinfo {pages} {161102} (\bibinfo {year} {2013})}\BibitemShut {NoStop}%
\bibitem [{\citenamefont {Casperson}\ and\ \citenamefont
  {Linn}(2006)}]{Casperson:VisualizationTeaching}%
  \BibitemOpen
  \bibfield  {author} {\bibinfo {author} {\bibfnamefont {Janet~M.}\
  \bibnamefont {Casperson}}\ and\ \bibinfo {author} {\bibfnamefont {Marcia~C.}\
  \bibnamefont {Linn}},\ }\bibfield  {title} {\enquote {\bibinfo {title}
  {{Using visualizations to teach electrostatics}},}\ }\href {\doibase
  10.1119/1.2186335} {\bibfield  {journal} {\bibinfo  {journal} {Am. J. Phys.}\
  }\textbf {\bibinfo {volume} {74}},\ \bibinfo {pages} {316--323} (\bibinfo
  {year} {2006})}\BibitemShut {NoStop}%
\bibitem [{\citenamefont {Tobochnik}\ and\ \citenamefont
  {Gould}(2008)}]{Tobochnik:StatPhysTeach}%
  \BibitemOpen
  \bibfield  {author} {\bibinfo {author} {\bibfnamefont {Jan}\ \bibnamefont
  {Tobochnik}}\ and\ \bibinfo {author} {\bibfnamefont {Harvey}\ \bibnamefont
  {Gould}},\ }\bibfield  {title} {\enquote {\bibinfo {title} {{Teaching
  statistical physics by thinking about models and algorithms}},}\ }\href
  {\doibase 10.1119/1.2839094} {\bibfield  {journal} {\bibinfo  {journal} {Am.
  J. Phys.}\ }\textbf {\bibinfo {volume} {76}},\ \bibinfo {pages} {353}
  (\bibinfo {year} {2008})}\BibitemShut {NoStop}%
\bibitem [{\citenamefont {Wieman}\ \emph {et~al.}(2008)\citenamefont {Wieman},
  \citenamefont {Perkins},\ and\ \citenamefont {Adams}}]{Wieman:PhetSoftware}%
  \BibitemOpen
  \bibfield  {author} {\bibinfo {author} {\bibfnamefont {Carl~E.}\ \bibnamefont
  {Wieman}}, \bibinfo {author} {\bibfnamefont {Katherine~K.}\ \bibnamefont
  {Perkins}}, \ and\ \bibinfo {author} {\bibfnamefont {Wendy~K.}\ \bibnamefont
  {Adams}},\ }\bibfield  {title} {\enquote {\bibinfo {title} {{Oersted Medal
  Lecture 2007: Interactive simulations for teaching physics: What works, what
  doesn't, and why}},}\ }\href {\doibase 10.1119/1.2815365} {\bibfield
  {journal} {\bibinfo  {journal} {Am. J. Phys.}\ }\textbf {\bibinfo {volume}
  {76}},\ \bibinfo {pages} {393} (\bibinfo {year} {2008})}\BibitemShut
  {NoStop}%
\bibitem [{\citenamefont {Laverty}\ and\ \citenamefont
  {Kortemeyer}(2012)}]{Laverty:Teaching}%
  \BibitemOpen
  \bibfield  {author} {\bibinfo {author} {\bibfnamefont {James}\ \bibnamefont
  {Laverty}}\ and\ \bibinfo {author} {\bibfnamefont {Gerd}\ \bibnamefont
  {Kortemeyer}},\ }\bibfield  {title} {\enquote {\bibinfo {title} {{Function
  plot response: A scalable system for teaching kinematics graphs}},}\ }\href
  {\doibase 10.1119/1.4719112} {\bibfield  {journal} {\bibinfo  {journal} {Am.
  J. Phys.}\ }\textbf {\bibinfo {volume} {80}},\ \bibinfo {pages} {724}
  (\bibinfo {year} {2012})}\BibitemShut {NoStop}%
\bibitem [{\citenamefont {Buffler}\ \emph {et~al.}(2008)\citenamefont
  {Buffler}, \citenamefont {Pillay}, \citenamefont {Lubben},\ and\
  \citenamefont {Fearick}}]{Buffler:ModelBasedTeaching}%
  \BibitemOpen
  \bibfield  {author} {\bibinfo {author} {\bibfnamefont {Andy}\ \bibnamefont
  {Buffler}}, \bibinfo {author} {\bibfnamefont {Seshini}\ \bibnamefont
  {Pillay}}, \bibinfo {author} {\bibfnamefont {Fred}\ \bibnamefont {Lubben}}, \
  and\ \bibinfo {author} {\bibfnamefont {Roger}\ \bibnamefont {Fearick}},\
  }\bibfield  {title} {\enquote {\bibinfo {title} {{A model-based view of
  physics for computational activities in the introductory physics course}},}\
  }\href {\doibase 10.1119/1.2835045} {\bibfield  {journal} {\bibinfo
  {journal} {Am. J. Phys.}\ }\textbf {\bibinfo {volume} {76}},\ \bibinfo
  {pages} {431--437} (\bibinfo {year} {2008})}\BibitemShut {NoStop}%
\bibitem [{\citenamefont {Henderson}(1974)}]{henderson:uniqueg}%
  \BibitemOpen
  \bibfield  {author} {\bibinfo {author} {\bibfnamefont {R.L.}\ \bibnamefont
  {Henderson}},\ }\bibfield  {title} {\enquote {\bibinfo {title} {A uniqueness
  theorem for fluid pair correlation functions},}\ }\href {\doibase
  http://dx.doi.org/10.1016/0375-9601(74)90847-0} {\bibfield  {journal}
  {\bibinfo  {journal} {Physics Letters A}\ }\textbf {\bibinfo {volume} {49}},\
  \bibinfo {pages} {197 -- 198} (\bibinfo {year} {1974})}\BibitemShut {NoStop}%
\bibitem [{\citenamefont {Davoudi}\ \emph {et~al.}(2000)\citenamefont
  {Davoudi}, \citenamefont {Kohandel}, \citenamefont {Mohammadi},\ and\
  \citenamefont {Tanatar}}]{davoudi:yukawa}%
  \BibitemOpen
  \bibfield  {author} {\bibinfo {author} {\bibfnamefont {B.}~\bibnamefont
  {Davoudi}}, \bibinfo {author} {\bibfnamefont {M.}~\bibnamefont {Kohandel}},
  \bibinfo {author} {\bibfnamefont {M.}~\bibnamefont {Mohammadi}}, \ and\
  \bibinfo {author} {\bibfnamefont {B.}~\bibnamefont {Tanatar}},\ }\bibfield
  {title} {\enquote {\bibinfo {title} {Hard-core {Y}ukawa model for
  charge-stabilized colloids},}\ }\href {\doibase 10.1103/PhysRevE.62.6977}
  {\bibfield  {journal} {\bibinfo  {journal} {Phys. Rev. E}\ }\textbf {\bibinfo
  {volume} {62}},\ \bibinfo {pages} {6977--6981} (\bibinfo {year}
  {2000})}\BibitemShut {NoStop}%
\bibitem [{\citenamefont {Cochran}\ and\ \citenamefont
  {Chiew}(2004)}]{cochran:yukawa}%
  \BibitemOpen
  \bibfield  {author} {\bibinfo {author} {\bibfnamefont {T.~W.}\ \bibnamefont
  {Cochran}}\ and\ \bibinfo {author} {\bibfnamefont {Y.~C.}\ \bibnamefont
  {Chiew}},\ }\bibfield  {title} {\enquote {\bibinfo {title} {Thermodynamic and
  structural properties of repulsive hard-core {Y}ukawa fluid: Integral
  equation theory, perturbation theory and {M}onte {C}arlo simulations},}\
  }\href {\doibase 10.1063/1.1759616} {\bibfield  {journal} {\bibinfo
  {journal} {J. Chem. Phys.}\ }\textbf {\bibinfo {volume} {121}},\ \bibinfo
  {pages} {1480--1486} (\bibinfo {year} {2004})}\BibitemShut {NoStop}%
\bibitem [{\citenamefont {Asakura}\ and\ \citenamefont
  {Oosawa}(1958)}]{asakura:asakuraoosawa}%
  \BibitemOpen
  \bibfield  {author} {\bibinfo {author} {\bibfnamefont {Sho}\ \bibnamefont
  {Asakura}}\ and\ \bibinfo {author} {\bibfnamefont {Fumio}\ \bibnamefont
  {Oosawa}},\ }\bibfield  {title} {\enquote {\bibinfo {title} {Interaction
  between particles suspended in solutions of macromolecules},}\ }\href
  {\doibase 10.1002/pol.1958.1203312618} {\bibfield  {journal} {\bibinfo
  {journal} {J. Polym. Sci.}\ }\textbf {\bibinfo {volume} {33}},\ \bibinfo
  {pages} {183--192} (\bibinfo {year} {1958})}\BibitemShut {NoStop}%
\bibitem [{\citenamefont {Roth}\ \emph {et~al.}(2000)\citenamefont {Roth},
  \citenamefont {Evans},\ and\ \citenamefont {Dietrich}}]{roth:asakuraoosawa}%
  \BibitemOpen
  \bibfield  {author} {\bibinfo {author} {\bibfnamefont {R.}~\bibnamefont
  {Roth}}, \bibinfo {author} {\bibfnamefont {R.}~\bibnamefont {Evans}}, \ and\
  \bibinfo {author} {\bibfnamefont {S.}~\bibnamefont {Dietrich}},\ }\bibfield
  {title} {\enquote {\bibinfo {title} {Depletion potential in hard-sphere
  mixtures: Theory and applications},}\ }\href {\doibase
  10.1103/PhysRevE.62.5360} {\bibfield  {journal} {\bibinfo  {journal} {Phys.
  Rev. E}\ }\textbf {\bibinfo {volume} {62}},\ \bibinfo {pages} {5360--5377}
  (\bibinfo {year} {2000})}\BibitemShut {NoStop}%
\bibitem [{\citenamefont {Hlushak}\ \emph {et~al.}(2013)\citenamefont
  {Hlushak}, \citenamefont {Hlushak},\ and\ \citenamefont
  {Trokhymchuk}}]{Hlushak:ModifiedFMSA}%
  \BibitemOpen
  \bibfield  {author} {\bibinfo {author} {\bibfnamefont {S.~P.}\ \bibnamefont
  {Hlushak}}, \bibinfo {author} {\bibfnamefont {P.~A.}\ \bibnamefont
  {Hlushak}}, \ and\ \bibinfo {author} {\bibfnamefont {A.}~\bibnamefont
  {Trokhymchuk}},\ }\bibfield  {title} {\enquote {\bibinfo {title} {{An
  improved first-order mean spherical approximation theory for the
  square-shoulder fluid.}}}\ }\href {\doibase 10.1063/1.4801659} {\bibfield
  {journal} {\bibinfo  {journal} {J. Chem. Phys.}\ }\textbf {\bibinfo {volume}
  {138}},\ \bibinfo {pages} {164107} (\bibinfo {year} {2013})}\BibitemShut
  {NoStop}%
\bibitem [{\citenamefont {Bannerman}\ and\ \citenamefont
  {Lue}(2010)}]{bannerman:pressure}%
  \BibitemOpen
  \bibfield  {author} {\bibinfo {author} {\bibfnamefont {Marcus~N.}\
  \bibnamefont {Bannerman}}\ and\ \bibinfo {author} {\bibfnamefont {Leo}\
  \bibnamefont {Lue}},\ }\bibfield  {title} {\enquote {\bibinfo {title} {Exact
  on-event expressions for discrete potential systems},}\ }\href {\doibase
  http://dx.doi.org/10.1063/1.3486567} {\bibfield  {journal} {\bibinfo
  {journal} {The Journal of Chemical Physics}\ }\textbf {\bibinfo {volume}
  {133}},\ \bibinfo {eid} {124506} (\bibinfo {year} {2010})}\BibitemShut
  {NoStop}%
\bibitem [{\citenamefont {Carmer}\ \emph {et~al.}(2012)\citenamefont {Carmer},
  \citenamefont {Goel}, \citenamefont {Pond}, \citenamefont {Errington},\ and\
  \citenamefont {Truskett}}]{carmer:s2opt}%
  \BibitemOpen
  \bibfield  {author} {\bibinfo {author} {\bibfnamefont {James}\ \bibnamefont
  {Carmer}}, \bibinfo {author} {\bibfnamefont {Gaurav}\ \bibnamefont {Goel}},
  \bibinfo {author} {\bibfnamefont {Mark~J.}\ \bibnamefont {Pond}}, \bibinfo
  {author} {\bibfnamefont {Jeffrey~R.}\ \bibnamefont {Errington}}, \ and\
  \bibinfo {author} {\bibfnamefont {Thomas~M.}\ \bibnamefont {Truskett}},\
  }\bibfield  {title} {\enquote {\bibinfo {title} {Enhancing tracer diffusivity
  by tuning interparticle interactions and coordination shell structure},}\
  }\href {\doibase 10.1039/C1SM06932B} {\bibfield  {journal} {\bibinfo
  {journal} {Soft Matter}\ }\textbf {\bibinfo {volume} {8}},\ \bibinfo {pages}
  {4083--4089} (\bibinfo {year} {2012})}\BibitemShut {NoStop}%
\bibitem [{\citenamefont {Pond}\ \emph {et~al.}(2009)\citenamefont {Pond},
  \citenamefont {Krekelberg}, \citenamefont {Shen}, \citenamefont {Errington},\
  and\ \citenamefont {Truskett}}]{pond:s2gauss}%
  \BibitemOpen
  \bibfield  {author} {\bibinfo {author} {\bibfnamefont {Mark~J.}\ \bibnamefont
  {Pond}}, \bibinfo {author} {\bibfnamefont {William~P.}\ \bibnamefont
  {Krekelberg}}, \bibinfo {author} {\bibfnamefont {Vincent~K.}\ \bibnamefont
  {Shen}}, \bibinfo {author} {\bibfnamefont {Jeffrey~R.}\ \bibnamefont
  {Errington}}, \ and\ \bibinfo {author} {\bibfnamefont {Thomas~M.}\
  \bibnamefont {Truskett}},\ }\bibfield  {title} {\enquote {\bibinfo {title}
  {Composition and concentration anomalies for structure and dynamics of
  {G}aussian-core mixtures},}\ }\href {\doibase
  http://dx.doi.org/10.1063/1.3256235} {\bibfield  {journal} {\bibinfo
  {journal} {J. Chem. Phys.}\ }\textbf {\bibinfo {volume} {131}},\ \bibinfo
  {eid} {161101} (\bibinfo {year} {2009})}\BibitemShut {NoStop}%
\bibitem [{\citenamefont {Krekelberg}\ \emph {et~al.}(2007)\citenamefont
  {Krekelberg}, \citenamefont {Mittal}, \citenamefont {Ganesan},\ and\
  \citenamefont {Truskett}}]{krekelberg:s2}%
  \BibitemOpen
  \bibfield  {author} {\bibinfo {author} {\bibfnamefont {William~P.}\
  \bibnamefont {Krekelberg}}, \bibinfo {author} {\bibfnamefont {Jeetain}\
  \bibnamefont {Mittal}}, \bibinfo {author} {\bibfnamefont {Venkat}\
  \bibnamefont {Ganesan}}, \ and\ \bibinfo {author} {\bibfnamefont {Thomas~M.}\
  \bibnamefont {Truskett}},\ }\bibfield  {title} {\enquote {\bibinfo {title}
  {How short-range attractions impact the structural order, self-diffusivity,
  and viscosity of a fluid},}\ }\href {\doibase
  http://dx.doi.org/10.1063/1.2753154} {\bibfield  {journal} {\bibinfo
  {journal} {J. Chem. Phys.}\ }\textbf {\bibinfo {volume} {127}},\ \bibinfo
  {eid} {044502} (\bibinfo {year} {2007})}\BibitemShut {NoStop}%
\bibitem [{\citenamefont {Rosenfeld}(1999)}]{rosenfeld:sexscaling1}%
  \BibitemOpen
  \bibfield  {author} {\bibinfo {author} {\bibfnamefont {Yaakov}\ \bibnamefont
  {Rosenfeld}},\ }\bibfield  {title} {\enquote {\bibinfo {title} {A
  quasi-universal scaling law for atomic transport in simple fluids},}\ }\href
  {http://stacks.iop.org/0953-8984/11/i=28/a=303} {\bibfield  {journal}
  {\bibinfo  {journal} {Journal of Physics: Condensed Matter}\ }\textbf
  {\bibinfo {volume} {11}},\ \bibinfo {pages} {5415} (\bibinfo {year}
  {1999})}\BibitemShut {NoStop}%
\bibitem [{\citenamefont {Rosenfeld}(1977)}]{rosenfeld:sexscaling2}%
  \BibitemOpen
  \bibfield  {author} {\bibinfo {author} {\bibfnamefont {Yaakov}\ \bibnamefont
  {Rosenfeld}},\ }\bibfield  {title} {\enquote {\bibinfo {title} {Relation
  between the transport coefficients and the internal entropy of simple
  systems},}\ }\href {\doibase 10.1103/PhysRevA.15.2545} {\bibfield  {journal}
  {\bibinfo  {journal} {Phys. Rev. A}\ }\textbf {\bibinfo {volume} {15}},\
  \bibinfo {pages} {2545--2549} (\bibinfo {year} {1977})}\BibitemShut {NoStop}%
\bibitem [{\citenamefont {Dzugutov}(1996)}]{dzugutov:sexscaling}%
  \BibitemOpen
  \bibfield  {author} {\bibinfo {author} {\bibfnamefont {Mikhail}\ \bibnamefont
  {Dzugutov}},\ }\bibfield  {title} {\enquote {\bibinfo {title} {A universal
  scaling law for atomic diffusion in condensed matter},}\ }\href {\doibase
  10.1038/381137a0} {\bibfield  {journal} {\bibinfo  {journal} {Nature}\
  }\textbf {\bibinfo {volume} {381}},\ \bibinfo {pages} {137--139} (\bibinfo
  {year} {1996})}\BibitemShut {NoStop}%
\bibitem [{\citenamefont {Oracle}()}]{swinglib}%
  \BibitemOpen
  \bibfield  {author} {\bibinfo {author} {\bibnamefont {Oracle}},\ }\href@noop
  {} {\emph {\bibinfo {title} {Java Swing Library}}}\BibitemShut {NoStop}%
\bibitem [{\citenamefont {{Object Refinery Limited}}(2013)}]{jfreechart}%
  \BibitemOpen
  \bibfield  {author} {\bibinfo {author} {\bibnamefont {{Object Refinery
  Limited}}},\ }\href {http://www.jfree.org/jfreechart/} {\emph {\bibinfo
  {title} {{JFreeChart} Version~1.0.17}}} (\bibinfo {year} {2013})\BibitemShut
  {NoStop}%
\bibitem [{\citenamefont {Jagla}(1999)}]{Jagla:ramp}%
  \BibitemOpen
  \bibfield  {author} {\bibinfo {author} {\bibfnamefont {E.~A.}\ \bibnamefont
  {Jagla}},\ }\bibfield  {title} {\enquote {\bibinfo {title} {Core-softened
  potentials and the anomalous properties of water},}\ }\href {\doibase
  http://dx.doi.org/10.1063/1.480241} {\bibfield  {journal} {\bibinfo
  {journal} {J. Chem. Phys.}\ }\textbf {\bibinfo {volume} {111}},\ \bibinfo
  {pages} {8980--8986} (\bibinfo {year} {1999})}\BibitemShut {NoStop}%
\bibitem [{\citenamefont {Yan}\ \emph {et~al.}(2006)\citenamefont {Yan},
  \citenamefont {Buldyrev}, \citenamefont {Giovambattista}, \citenamefont
  {Debenedetti},\ and\ \citenamefont {Stanley}}]{yan:ramp}%
  \BibitemOpen
  \bibfield  {author} {\bibinfo {author} {\bibfnamefont {Zhenyu}\ \bibnamefont
  {Yan}}, \bibinfo {author} {\bibfnamefont {Sergey~V.}\ \bibnamefont
  {Buldyrev}}, \bibinfo {author} {\bibfnamefont {Nicolas}\ \bibnamefont
  {Giovambattista}}, \bibinfo {author} {\bibfnamefont {Pablo~G.}\ \bibnamefont
  {Debenedetti}}, \ and\ \bibinfo {author} {\bibfnamefont {H.~Eugene}\
  \bibnamefont {Stanley}},\ }\bibfield  {title} {\enquote {\bibinfo {title}
  {Family of tunable spherically symmetric potentials that span the range from
  hard spheres to waterlike behavior},}\ }\href {\doibase
  10.1103/PhysRevE.73.051204} {\bibfield  {journal} {\bibinfo  {journal} {Phys.
  Rev. E}\ }\textbf {\bibinfo {volume} {73}},\ \bibinfo {pages} {051204}
  (\bibinfo {year} {2006})}\BibitemShut {NoStop}%
\bibitem [{\citenamefont {Errington}\ \emph {et~al.}(2006)\citenamefont
  {Errington}, \citenamefont {Truskett},\ and\ \citenamefont
  {Mittal}}]{errington:ramp}%
  \BibitemOpen
  \bibfield  {author} {\bibinfo {author} {\bibfnamefont {Jeffrey~R.}\
  \bibnamefont {Errington}}, \bibinfo {author} {\bibfnamefont {Thomas~M.}\
  \bibnamefont {Truskett}}, \ and\ \bibinfo {author} {\bibfnamefont {Jeetain}\
  \bibnamefont {Mittal}},\ }\bibfield  {title} {\enquote {\bibinfo {title}
  {Excess-entropy-based anomalies for a waterlike fluid},}\ }\href {\doibase
  10.1063/1.2409932} {\bibfield  {journal} {\bibinfo  {journal} {J. Chem.
  Phys.}\ }\textbf {\bibinfo {volume} {125}},\ \bibinfo {eid} {244502}
  (\bibinfo {year} {2006})}\BibitemShut {NoStop}%
\bibitem [{\citenamefont {Foundation}(2007)}]{gpl}%
  \BibitemOpen
  \bibfield  {author} {\bibinfo {author} {\bibfnamefont {Free~Software}\
  \bibnamefont {Foundation}},\ }\href {http://www.gnu.org/copyleft/gpl.html}
  {\emph {\bibinfo {title} {{GNU} General Public License}}} (\bibinfo {year}
  {2007})\BibitemShut {NoStop}%
\bibitem [{git()}]{github}%
  \BibitemOpen
  \href@noop {} {}\BibitemShut {NoStop}%
\bibitem [{\citenamefont {Hansen}\ and\ \citenamefont
  {Verlet}(1969)}]{hansen:freezing}%
  \BibitemOpen
  \bibfield  {author} {\bibinfo {author} {\bibfnamefont {Jean-Pierre}\
  \bibnamefont {Hansen}}\ and\ \bibinfo {author} {\bibfnamefont {Loup}\
  \bibnamefont {Verlet}},\ }\bibfield  {title} {\enquote {\bibinfo {title}
  {Phase transitions of the {L}ennard-{J}ones system},}\ }\href {\doibase
  10.1103/PhysRev.184.151} {\bibfield  {journal} {\bibinfo  {journal} {Phys.
  Rev.}\ }\textbf {\bibinfo {volume} {184}},\ \bibinfo {pages} {151--161}
  (\bibinfo {year} {1969})}\BibitemShut {NoStop}%
\end{thebibliography}
\end{document}